\documentclass[%
reprint,
superscriptaddress,
longbibliography,
amsmath,amssymb,
aps,
prl,
]{revtex4-1}

\usepackage{graphicx}
\usepackage{dcolumn}
\usepackage{bm}
\usepackage{hyperref}
\hypersetup{%
	colorlinks = true,
	citecolor = blue,
	linkcolor = blue,
	urlcolor = blue,
	linkbordercolor = 0 0 0,
	pdfborder= 0 0 0,
}

\usepackage[pagewise, mathlines]{lineno}
\usepackage{physics}
\usepackage{xcolor}

\usepackage{multirow}

\usepackage{gensymb}
\begin{document}

\title{Precision Positronium Spectroscopy as a Test of the Helium Ionization Energy Anomaly}


\author{J. P\'{e}rez-R\'{i}os}
\address{Department of Physics and Astronomy, Stony Brook University, Stony Brook, New York 11794, USA}

\author{D. B. Cassidy}
\affiliation{Department of Physics and Astronomy, University College London, Gower Street, London, WC1E 6BT, United Kingdom}

\date{\today}

\begin{abstract}
Precise measurements of the metastable helium (1s)(2s)$\,^3S_1$ ionization energy have revealed a 9$\sigma$ discrepancy with QED theory that persists in isotopic measurements, suggesting a leptophilic bosonic interaction as a possible explanation. A subsequent investigation of such interactions has concluded that only a scalar boson interaction is consistent with the He observations. Taking this as a starting point, we derive the response of positronium energy levels to the corresponding finite-range Yukawa potential, using exact hydrogenic matrix elements and a numerical helium calculation. Across the viable mediator-mass range of 0-800 eV, the He anomaly interpreted in this way implies a positronium $1\,^3$S$_1 \rightarrow 2\,^3$S$_1$ shift ranging from \(0.250\)--\(0.850\) MHz, and a \(2S\)-ionization shift of \(0.14\)--\(0.21\)~MHz. We discuss the feasibility of observing these shifts experimentally.

\end{abstract}


\maketitle


Precision spectroscopy of simple atomic systems provides some of the most stringent tests of bound-state quantum electrodynamics (QED) and continues to play a central role in the determination of fundamental constants and searches for physics beyond the Standard Model~\cite{Karshenboim2005}. Advances in both experiment and theory have produced remarkable agreement for few-body systems, especially hydrogen and helium~\cite{drake2023springer}. Consequently, a statistically significant discrepancy between a precision measurement and high-order theoretical calculations is of particular interest, since it may indicate either an incomplete treatment of conventional bound-state physics, or the presence of a previously unrecognized interaction.

Recently a high-precision determination of the metastable triplet $^3$He (1s)(2s)$\,^3S_1$ ionization energy~ was found to disagree significantly with the best available theoretical prediction: the experimental determination of the ionization frequency was $1,152,788,844.6154 \pm 0.0081$~MHz~\cite{Clausen2025b}, while the QED calculation finds $1,152,788,844.1334 \pm 0.0520$~MHz~\cite{Patkos2021}. The disagreement then amounts to $\Delta E^{\text{3He}} = -482.0 \pm 52.6$~kHz. This result confirmed a previous measurement made using $^4$He in which the ionization frequency obtained was $1,152,842,742.7082 \pm 0.0060$~MHz~\cite{Clausen2025}, which differs from the corresponding theory value of $1,152,842,742.2310 \pm 0.0520$~MHz~\cite{Patkos2021} by a similar amount, yielding $\Delta E^{\text{4He}} = -477.2 \pm 52.3$~kHz. Both of these correspond to $\approx 9 \sigma$ discrepancies, and it is notable that the uncertainty comes almost entirely from the theory side.

One proposed explanation for the observed discrepancy is a weak finite-range scalar interaction between charged leptons mediated by a new boson~\cite{Cong2026}. An interaction between leptons is indicated from the fact the two helium isotopes both exhibit almost identical anomalies. This, along with other direct measurements~\cite{Steinbach2026}, rules out finite nuclear size effects and electron-nucleon interactions~\cite{Karshenboim2010}, and hence no effect from the hypothesized interaction would be expected in hydrogen measurements, which have been performed with high precision (e.g., ~\cite{Fleurbaey2018, Bezginov2019, Scheidegger2024, Maisenbacher2026, Bullis2026}), including 1S-2S measurements, which have achieved extreme precision of $4.2 \times 10^{-15}$~\cite{Parthey2011}. 

The electron-positron bound state positronium (Ps) represents a complementary atomic system that is both hydrogenic, and contains two leptons~\cite{Wheeler1946}. This means that the Ps bound-state wavefunctions are known analytically, and its spectrum can be calculated with high precision~\cite{Adkins2022}. While helium derives much of its sensitivity from correlated electron-electron dynamics, Ps is a simple two-body system with no nuclear-structure contribution. The two systems are therefore complementary probes of the proposed leptonic interaction: if a scalar force is responsible for the helium anomaly, it must generate a correlated and quantitatively calculable pattern of shifts in positronium which can, in principle, be probed using precision Ps spectroscopy. To date, no quantitative framework has connected the scalar interpretation of the helium ionization-energy anomaly to experimentally testable positronium observables. In this work we establish such a framework by deriving analytic response functions for hydrogenic positronium states in a finite-range Yukawa potential. We then relate the positronium shifts to the helium discrepancy through a benchmarked correlated calculation of the metastable-helium matrix element. The resulting predictions identify the most sensitive measurements over the relevant mediator-mass range and clarify the complementary roles of $1\,^3$S$_1 \rightarrow 2\,^3$S$_1$ spectroscopy and $2\,^3$S$_1 \rightarrow$~ionization measurements. 


To investigate the implications of the hypothesized scalar field interaction for positronium, and following the analysis of Cong \textit{et al.}~\cite{Cong2026}, we assume that the helium anomaly originates from a finite-range scalar Yukawa interaction of the form
\begin{equation}
\label{Vs}
V(r)=-g_e^2\frac{\hbar c}{4\pi}\frac{e^{-r/\lambda}}{r},
\end{equation}
where $g_e$ is the coupling constant of the new scalar field to electrons and positrons, $\hbar$ is the reduced Planck constant, $c$ is the speed of light in vacuum, and $\lambda=\hbar/m_{\phi}c$ defines the interaction range associated with a mediator of mass $m_{\phi}$. The scalar interaction in Eq.~(\ref{Vs}) is expected to be much weaker than the Coulomb binding potential of positronium; its contribution to the energy levels can therefore be treated, to leading order, using first-order perturbation theory:
\begin{equation}
\Delta E_{n\ell}=
-g_e^2\frac{\hbar c}{4\pi}
F_{n\ell}(\lambda),
\label{eq:GeneralEnergyShift}
\end{equation}
where
\begin{equation}
F_{n\ell}(\lambda)=
\int_0^\infty
|R_{n\ell}(r)|^2
e^{-r/\lambda}
r,dr,
\label{eq:IntegralDefinition}
\end{equation}
and $R_{n\ell}(r)$ denotes the normalized hydrogenic radial wavefunction with principal quantum number $n$ and orbital angular momentum $\ell$. The quantity $F_{n\ell}$ contains the dependence of the atomic response on the interaction range, and the overall interaction strength is determined by the dimensionless scalar coupling parameter $g_e$.

Unlike helium, whose accurate description requires correlated two-electron wavefunctions, positronium is an exactly hydrogenic two-body system, with the electron-positron reduced mass incorporated into its Bohr radius (i.e., $a_{\rm Ps}=2a_0$, where $a_0$ is the Bohr radius~\cite{Bethe1957}). Consequently, the radial integrals can be evaluated analytically. For the states of interest here we obtain
\begin{eqnarray}
F_{1S}&=&\frac{4}{a_{\rm Ps}(2+x)^2}, \label{1s} \\
F_{2S}&=&\frac{1+2x^2}{4a_{\rm Ps}(1+x)^4},\label{2s} \\
F_{2P}&=&\frac{1}{4a_{\rm Ps}(1+x)^4},\label{2p}
\end{eqnarray}
where $x={a_{\rm Ps}} / \lambda$.

\begin{figure}[t]
    \centering
    \includegraphics[width=\linewidth]{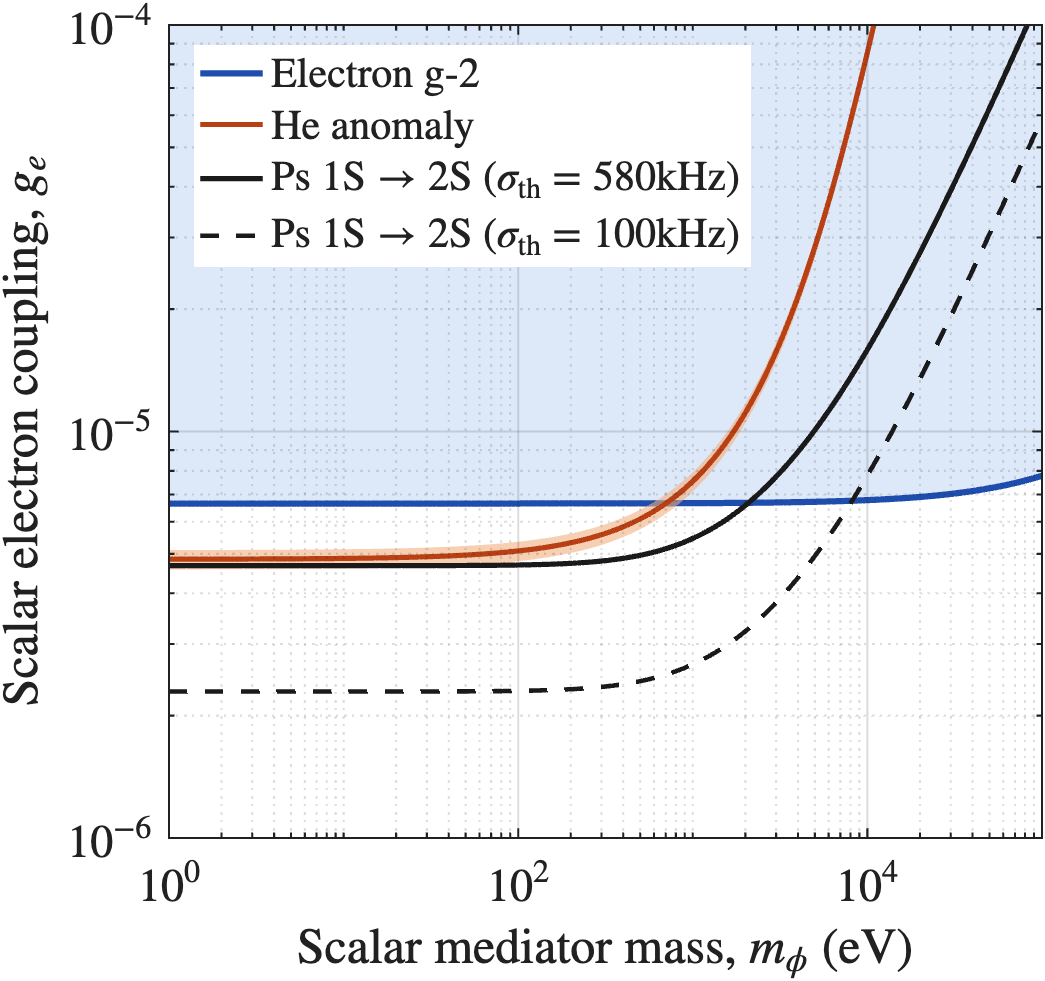}
    \caption{Scalar electron coupling $g_e$ mediated by a hypothetical boson of mass $m_\phi$. The blue shaded region shows the parameter space excluded by measurements of the electron anomalous magnetic moment, $g-2$~\cite{Frugiuele2019}. The orange shaded region indicates the coupling range consistent with the helium ionization-energy anomaly ($\Delta E^{\text{He}}=-477\pm52$~kHz, see text for details). The solid black curve represents the projected sensitivity of a Ps $1\,^3$S$_1 \rightarrow 2\,^3$S$_1$ measurement with an experimental precision of $100~\mathrm{kHz}$, including the current theoretical uncertainty of $580~\mathrm{kHz}$. The dashed black curve assumes the same experimental precision, but a reduced theoretical uncertainty of $100~\mathrm{kHz}$.}
\label{fig:exclusion}
\end{figure}

Focusing first on the Ps $1\,^3$S$_1 \rightarrow 2\,^3$S$_1$ transition, we determine the constraint on the scalar coupling $g_e$ as a function of the mediator mass from the scalar-induced shift of the transition energy, $\Delta E_{20}-\Delta E_{10}$. The resulting sensitivity is shown in Fig.~\ref{fig:exclusion}, together with the constraints derived from the helium anomaly and from measurements of the electron anomalous magnetic moment, (g-2)~\cite{Frugiuele2019}, through one loop corrections to the electron-photon interaction~\cite{Delaunay2017,NA64}. Positronium probes part of the parameter space in which the helium anomaly provides a stronger constraint than electron (g-2) measurements. The solid black curve in Fig.~\ref{fig:exclusion} shows the projected Ps sensitivity for an experimental uncertainty $\sigma_{\rm exp} = 100$~kHz, including the present theoretical uncertainty $\sigma_{\rm th} = 0.58$~MHz~\cite{Czarnecki1999}; these are added in quadrature. Under these assumptions, the projected Ps constraints are comparable to those arising from the helium anomaly and remain competitive even up to slightly larger mediator masses.


Work is in progress to calculate higher order terms for Ps energy levels~\cite{Adkins2018, Adkins:2019Hc}.  Simple power and logarithmic counting suggests that a theoretical uncertainty at the level of  $\sigma_{\rm th} \approx 100$~kHz may be achievable. If this could be obtained the corresponding projected sensitivity would be substantially improved, as indicated in Fig.~\ref{fig:exclusion} by the black dashed curve. In this scenario, a Ps measurement (again with $\sigma_{\rm exp} = 100$~kHz) would provide a stronger constraint than that associated with the helium anomaly, and would extend the sensitivity to mediator masses of order $10~{\rm keV}$. The general behavior of the Ps sensitivity can be understood from the different spatial and short-range properties of the Ps and He wavefunctions. Positronium is more spatially extended: since $\text{a}_{\rm Ps}=2\text{a}_0$, the Ps 1S wavefunction decays as $\exp[-r/(2a_0)]$, whereas the most diffuse component of the approximate metastable-He wavefunction used here decays as $\exp(-0.765\,r/a_0)$~\cite{Ficek2017}. This greater spatial extent makes Ps particularly sensitive to longer-range interactions and therefore to lighter mediators.

\begin{figure}[t]
    \centering
        \includegraphics[width=1\linewidth]{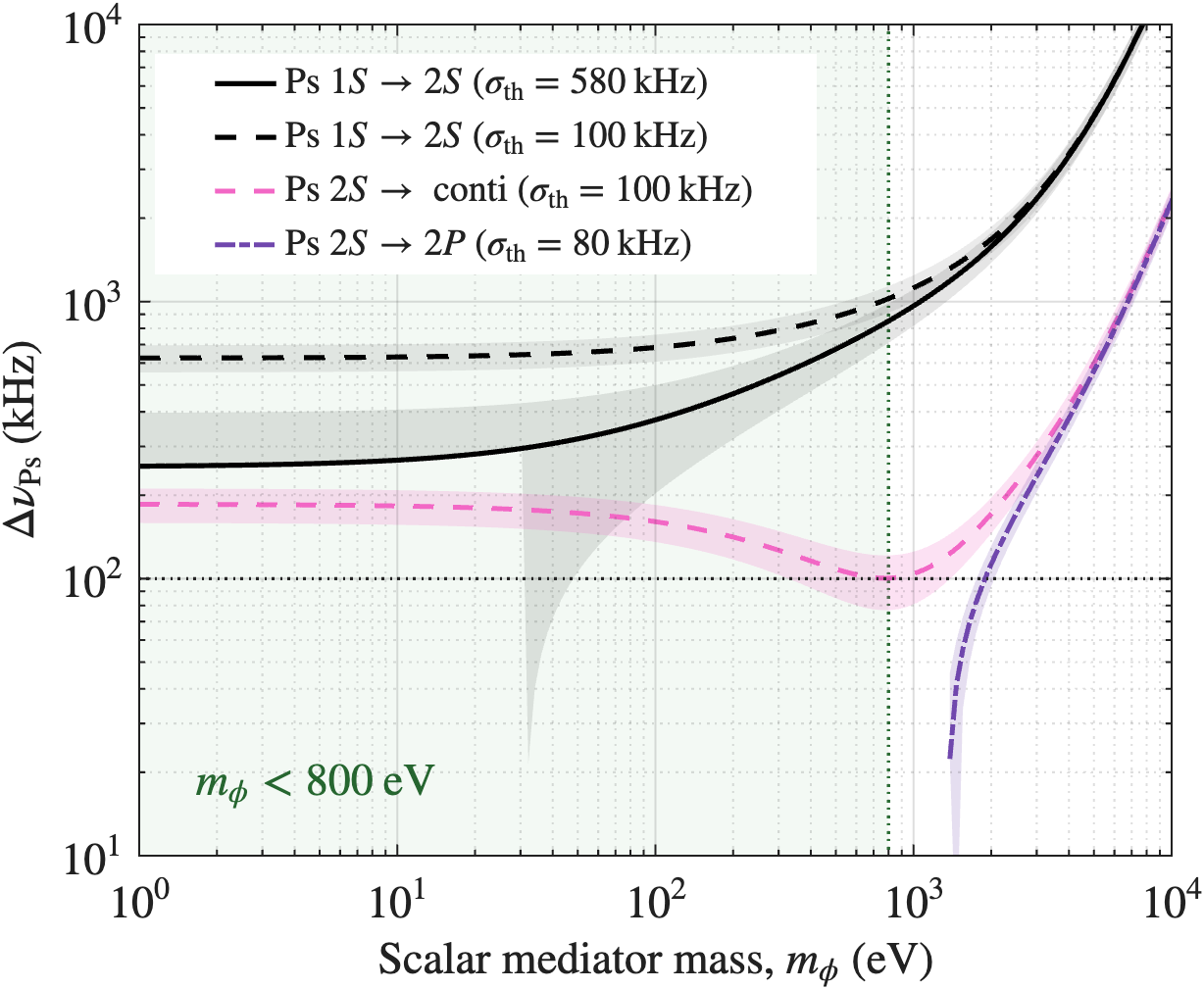}
    \caption{Frequency shift of various Ps transitions as described by Eq.~(\ref{eqn:freq_shift}) as a function of the scalar-mediator mass. The shaded bands represent the uncertainty propagated from the $\Delta E^{\text{He}}$ uncertainty, including the QED theoretical contribution as indicated in the legend.
    The uncertainty associated with the helium anomaly is propagated through the quadrature relation independently for its upper and lower bounds (see text for details).}
\label{fig:shift}
\end{figure}

In the case of Helium, the same scalar field will create an energy shift of the ionization energy of the 2$^3$S state, given by  
\begin{equation}
\Delta E^{\text{He}}
=
-g_e^2\frac{\hbar c}{4\pi} C_{\rm He}(\lambda), 
\label{eq:RadialIntegral}
\end{equation}
where
\begin{equation}
C_{\rm He}(\lambda)=
\left\langle
\frac{e^{-r_{12}/\lambda}}{r_{12}}
\right\rangle_{2\,^3S},
\end{equation}
and $r_{12}$ represents the electron-electron separation. Here $C_{\rm He}(\lambda)$ is evaluated numerically using the He wave function of Ref~\cite{Ficek2017}. Including the helium anomaly measurement, such that $\Delta E^{\text{He}}=-477\pm52$~kHz, it is possible to compute a corresponding energy shift in Ps transitions given by 
\begin{equation}
\Delta \nu _{\mathrm{Ps}}=\frac{\Delta E^{\text{He}}}{h}\frac{F_{n\ell}(\lambda)}{C_{\rm He}(\lambda)}.
\label{eqn:freq_shift}
\end{equation}
This shift has been evaluated for three different transitions in Ps, and the results are shown in Fig.~\ref{fig:shift}. The figure focuses on the mediator mass range where the helium anomaly could potentially be explained by a new scalar field (i.e., $m_{\phi} < 800$~eV). From the figure, it is clear that the $1\,^3$S$_1 \rightarrow 2\,^3$S$_1$ transition in Ps has the largest energy shift and therefore represents the most promising route to explore the parameter space describing the helium anomaly, although for light mediator masses the energy shift 2$^3$S$_1$ ionization becomes comparable. It can be seen in Fig.~\ref{fig:shift} that the uncertainty band for the $1S-2S$ transition including the current 580~kHz theory uncertainty is partially truncated. This occurs because, when one of the propagated bounds falls below the theoretical uncertainty, the scalar field induced energy shift cannot be unambiguously separated from our incomplete knowledge of the calculated Ps energy levels; if the theory uncertainty is reduced to 100~kHz then the energy shift is clearly defined. This highlights the urgent need not only for improved experimental precision, but also for new QED calculations.


The calculations presented above establish the scalar-induced frequency shifts expected in experimentally accessible positronium observables if the scalar interaction inferred from the reported helium ionization-energy anomaly is indeed realized in nature~\cite{Clausen2025b,Cong2026}. Here we consider how these predicted frequency shifts compare with realistic precision positronium spectroscopy.

The observables considered here differ in both their predicted scalar-induced frequency shifts and their current experimental status. However, rather than providing competing measurements, they fulfill complementary roles. Measurements of the Ps $1\,^3$S$_1 \rightarrow 2\,^3$S$_1$ interval are well established~\cite{Chu1982, Chu1984, Fee1993, Fee1993b}, and offer the greatest sensitivity to the existence of the proposed scalar interaction owing to the comparatively large predicted frequency shifts. In contrast, the $2\,^3$S$_1$ ionization-energy measurement has not been performed with high precision, but nevertheless represents a valuable addition as it probes different combinations of positronium states, and therefore could provide an independent test of the radial dependence of the interaction. In contrast, the $2^3S_1\rightarrow2^3P_J$ (fine structure) transitions are insensitive to very light scalar mediators. In the limit of vanishing mediator mass, ($x\rightarrow0$), Eqs.~(\ref{2s}) and~(\ref{2p}) become identical, so the scalar-induced shifts of the two levels cancel and the transition loses sensitivity to the new interaction. Hence, once the corresponding theory error ($\sigma_{\rm th} = 80$~kHz~\cite{Czarnecki1999}) is taken into account, there is essentially no observable $2S-2P$ shift in the low mediator mass range (see Fig.~\ref{fig:shift}).  

As shown in Fig.~\ref{fig:shift}, the predicted scalar-induced frequency shift of the Ps $1\,^3$S$_1 \rightarrow 2\,^3$S$_1$ transition lies in the range 0.250–0.850 MHz over the viable scalar-mass range. The experimental precision required to probe this effect is strongly limited by the current theoretical QED uncertainty of 580 kHz, as already indicated by the analysis of Fig.~\ref{fig:exclusion}. A next-generation QED calculation could reduce the theoretical uncertainty to approximately 100 kHz. In that case, the experimental uncertainty required for Ps to remain competitive would lie in the range 0.630–1.030 MHz. The larger allowed experimental uncertainty reflects the reduced theoretical contribution to the total uncertainty budget. The most precise measurement of the positronium $1\,^3$S$_1 \rightarrow 2\,^3$S$_1$ interval is $\nu_{\mathrm{1s-2s}}^{\mathrm{\exp}} = 1\,233\,607\,216.4 \pm 3.2$~MHz, measured by Fee \textit{et al.} in 1993~\cite{Fee1993b}. The QED Theory Prediction is $\nu_{\mathrm{1s-2s}}^{\mathrm{th}} = 1\,233\,607\,222.12 \pm 0.58$~MHz~\cite{Czarnecki1999}. There is a small (5.7~MHz) disagreement, amounting to $\approx 2 \sigma$, which is reduced slightly when the more recent (but less precise) result of Borges \textit{et al.} is included~\cite{Borges2026}. The measurement of Fee \textit{et al.} has stood since 1993 but it is expected that various technological advances such as more stable lasers, precise metrology from frequency combs, and positron beam and trap developments, will contribute to new experimental work aimed at reaching a precision of $\approx 100$~kHz~\cite{Cassidy2018}. Several new experimental programs have been discussed in the literature (e.g.,~\cite{Cooke2015, Mills2016x}), including using laser-cooled Ps~\cite{Gloggler2024, Shu2024}. Some new results have been produced~\cite{Borges2026}, and, although the Fee precision has not yet been surpassed, the application of new methods is expected to yield advances in the near future~\cite{Heiss2025, Javary2025}. 

The experiment by Clausen~\textit{et al.} that has motivated this work~\cite{Clausen2025} measured the triplet $^4$He ionization energy via the hyperfine resolved (1s)(2s)$^3S_1 \rightarrow np$ Rydberg series, with the principal quantum number $n$ ranging from 24 to 100~\cite{Clausen2025}. Experiments of this type with He are conducted using the metastable excited $2S$ states for practical reasons: starting from the ground state and driving the (1s)$^2 \rightarrow np$ series would require laser wavelengths in the extreme ultra-violet region (i.e., $\lambda \approx 50$~nm), as opposed to the much more accessible $\lambda \approx 260$~nm used in the experiments. To do the same measurement with positronium, starting from the $1\,^3$S$_1$ ground state, would require laser light with wavelengths $\lambda \approx 180$~nm. While this is not impossible to generate (e.g., \cite{HANNA2009}), it would still be preferable to perform a Ps ionization energy measurement via the  $2\,^3S_1 \rightarrow np$ Rydberg series because (a) the $2\,^3S_1$ lifetime is 8 times longer than that of the $1\,^3S_1$ ground state~\cite{Adkins2022}, meaning that the natural width of the transitions would be narrower (i.e., 140~kHz instead of 1.12~MHz) and (b) the required $\lambda \approx 730$~nm laser light is considerably easier to produce with high intensity.

Ps Rydberg series' have been previously observed~\cite{Ziock1990b, Cassidy2012, Wall2015, Aghion2016, Baker2018}. However, as these experiments were all performed using pulsed lasers, they were not high precision measurements. A significant complication related to performing a Ps analogue of the Clausen He measurements is that, as a hydrogenic system, Ps is extremely susceptible to $\ell$ mixing from very small stray electric fields, and in general one has to work in the parabolic Stark-mixed regime~\cite{Gallagher1994} in which individual Rydberg-Stark ($k$) states are addressed~\cite{Wall2015}. This means that the approach of Clausen \textit{et al.} would have to be modified to also deal with the effects of stray electric fields. This could be done using techniques similar to those described by Scheidegger \textit{et al.}~\cite{Scheidegger2023}, who were able to measure the energies of hydrogen Rydberg states with high precision~\cite{Scheidegger2024} by using the different sensitivities of various $k$ states to measure and compensate for stray electric fields. It should be pointed out that this would represent a very challenging and completely new experimental Ps programme, in contrast to $1\,^3$S$_1 \rightarrow 2\,^3$S$_1$ measurements, which are far more developed.    

Precision microwave spectroscopy of the positronium fine structure has advanced considerably in recent years, with measurements of the $2^3S_1\rightarrow2^3P_J$ intervals now reaching sub megahertz precision \cite{Gurung2020c,Sheldon2023b,Newson2026}. However, as is evident from Fig.~\ref{fig:shift}, there is no  associated shift for this transition over the maximum viable mediator mass. Therefore, we conclude that Ps $n=2$ fine structure measurements do not present an experimentally viable option for testing the scalar boson hypothesis.


If the helium discrepancy is due to a scalar interaction, the predicted positronium shifts are large enough to motivate dedicated measurements. We identify $1\,^3$S$_1 \rightarrow 2\,^3$S$_1$ spectroscopy as the most sensitive test of the leptonic interaction hypothesis, and Rydberg-ionization and Stark-manifold measurements as providing independent constraints on the excited-state shift and radial structure of the interaction. As indicated in Fig.~\ref{fig:shift}, an improved QED theory uncertainty would significantly increase the sensitivity of the $1\,^3$S$_1 \rightarrow 2\,^3$S$_1$ measurements, and in fact is required if the full mediator mass range is to be tested. A null positronium result would strongly disfavor the scalar boson interpretation, while correlated shifts in $1\,^3$S$_1 \rightarrow 2\,^3$S$_1$ and the $2\,^3$S$_1$~binding energy would test both the strength and radial form of the interaction.

Astrophysical energy-loss arguments can place substantially stronger bounds on light scalars coupled to electrons than the laboratory constraints considered here~\cite{Frugiuele2019}. Such bounds depend on the production, propagation, and screening properties of the mediator and therefore introduce additional model assumptions. The present analysis is restricted to the laboratory-compatible scalar parameter space identified in
Ref.~\cite{Cong2026} that can be directly tested experimentally. 

The unresolved helium ionization energy discrepancy is the immediate scientific driver of this work, but the positronium spectroscopy program we describe nevertheless has broader value: even if the helium anomaly is ultimately resolved within conventional QED theory, for example from an unknown theoretical contribution to the Lamb shifts of the 2$^3$S and 2$^3$P states of He~\cite{Patkos2021}, $1\,^3$S$_1 \rightarrow 2\,^3$S$_1$ measurements will continue to constitute much needed tests of bound-state QED~\cite{Adkins2022}, and the Rydberg measurement would provide a new absolute excited-state binding-energy determination in positronium.

\section*{Acknowledgments} 
This work was supported in part by the EPSRC under Grant EP/W032023/1. DBC is grateful to E. A. Cornell and S. D. Hogan for helpful discussions. J.P.-R. acknowledges the support of the Simons Foundation.

\end{document}